\documentstyle[grlga,graphicx]{article}

\lefthead{SORNETTE AND PISARENKO}
\righthead{FRACTAL PLATE TECTONICS}
\received{XXX}
\revised{XXX}
\accepted{XXX}
\paperid{}
\cpright{AGU}{2002}
\authoraddr{D. Sornette,
Laboratoire de Physique de la Mati\`{e}re Condens\'{e}e, CNRS UMR 6622 and
Universit\'{e} de Nice-Sophia Antipolis, Parc Valrose, 06108 Nice, France,
and Department of Earth and Space Sciences and Institute of
Geophysics and Planetary Physics, University of California, Los Angeles,
California 90095-1567.
(email: sornette@unice.fr)}

\authoraddr{V.F. Pisarenko,
International Institute of Earthquake Prediction Theory and
Mathematical Geophysics, Russian Ac. Sci. Warshavskoye sh., 79, kor. 2,
Moscow 113556, Russia.
(email: vlad@sirius.mitp.ru)}

\newcommand{\be}{\begin{equation}}
\newcommand{\ee}{\end{equation}}
\newcommand{\ba}{\begin{eqnarray}}
\newcommand{\ea}{\end{eqnarray}}

\begin{document}

%% ------------------------------------------------------ %%
%
%%  TITLE
%
%% ------------------------------------------------------ %%

\title{Fractal Plate Tectonics}

%% ------------------------------------------------------ %%
%
%%  AUTHOR NAMES, AFFILIATIONS, and ALTERNATE AFFILIATIONS
%
%% ------------------------------------------------------ %%

\author{Didier Sornette}
\affil{Laboratoire de Physique de la Mati\`{e}re Condens\'{e}e, CNRS UMR 6622 and
Universit\'{e} de Nice-Sophia Antipolis, Nice, France and
Department of Earth and Space Sciences and Institute of
Geophysics and Planetary Physics, University of California, Los Angeles, CA}

\author{Vladilen Pisarenko}
\affil{International Institute of Earthquake Prediction Theory and
Mathematical Geophysics, Russian Ac. Sci. Warshavskoye sh., 79, kor. 2,
Moscow 113556, Russia}

\begin{abstract}
We analyze in details the statistical significance of the claim by Bird [2002]
of a power law distribution of plate areas covering the Earth and confirm
that the power law with exponent $0.25 \pm 0.05$ 
is the most robust and parsimonious model for all plates, 
including the very largest plates, when taking
into account the constraint that the plates areas must sum up to $4 \pi$
steradians. We propose a general class
of fragmentation models that rationalize this observation and discuss
the implications for the earth dynamics and the
general self-organization of tectonic deformations at multiple scales.

\end{abstract}

   % The abstract consists of a one-paragraph summary
   % of your paper, 250 words or fewer.  Do not cite
   % references unless absolutely necessary.  If you
   % must cite, place the citation in square brackets
   % using italic type (this abstract citation format
   % is new).  Do not include displayed material.

%%%% LEVEL 4 HEADS
   %
   % Use the command \subsubsubsection{} to identify
   % Level 4 heads; type the appropriate head between
   % the curly brackets, as shown.
   %
   % Capitalize only acronyms, the first letter of the
   % first word, first letter of proper nouns, and first
   % letter of the first word after colons.
   %
   % Hyphenation is permitted in Level 4 heads, if needed.
   %
   % A level 4 head cannot directly follow a Level
   % 3 head; there must be a least one sentence between
   % the two heads.

\begin{article}
\section{Introduction}

According to the theory of plate tectonics, the outermost $100$
km of the solid Earth is composed of a relatively small number of internally
rigid plates (lithosphere) that move over a weak substrate (asthenosphere).
The plate sizes and positions change over time and are driven
by the internal mantle convection. DeMets et al. [1990] performed a global
inversion to determine the relative rotation rates of the 12 largest plates (the
NUVEL-1 model) later refined into the NUVEL-1A
solution [{\it DeMets et al.}, 1994]. The edges of these plates, where they move
against each other, are sites of intense geologic activity, such as earthquakes,
volcanoes, and mountain building. 
These plate edges are not sharp narrow boundaries but are often
constituted of complex systems of competing faults and other geological
structures of width extending over
several hundreds of kilometers for transform and subduction boundaries
and up to thousands kilometers for continental collisions. It is now common lore
to view such tectonic deformation
as possessing some kind of self-similarity or fractal properties
[{\it Scholz and Mandelbrot}, 1989], 
or better a hierarchical structure [{\it Ouillon et al.}, 1996].
In addition, several researchers have repeatedly proposed to describe this
multi-scale organization of faulting by a hierarchy of blocks of multiple sizes
[{\it Gelfand et al.}, 1976; {\it Gorshkov et al.}, 2001]. 
In these models, blocks slide against each other, 
rotate, lock at nodes or triple junctions which may represent the 
loci of major earthquakes,
in a way similar to
(but at a reduced scale compared to) the relative motion of the major tectonic plates.
An example of a complex network of faults is observed in the broad
San Andreas transform boundary between the Pacific and the North
American plates in California, for which Bird and Rosenstock [1984] have suggested
$21$ possible microplates within southern California alone.

Keeping in mind these ingredients of a few major plates at large scales on one hand and
a hierarchical self-similar organization of blocks at the boundary scale
on the other hand, the recent reassessment of 
present plate boundaries on the Earth by Bird [2002] is particularly 
inspiring: taking into account
relative plate velocities from magnetic anomalies, moment tensor solutions,
and/or geodesy, to the 14 large plates whose motion was described by
the NUVEL-1A poles, model PB2001 [Bird, 2002] includes 28 
additional small plates, for a total of 42 plates. Bird [2002] suggests
that the cumulative-number/area distribution for his model follows a
power-law for plates of less than 1 steradian of area.

\section{Statistical analysis of the size distribution of the $42$ plates}

\subsection{Analysis with rank-order statistics and the Pareto law}

The statistical analysis of such a data set is very difficult due to its small size $N=42$
but is not impossible. Figure 1 shows
the complementary cumulative number $N(A)$ of plates as a function of area $A$ in steradians,
i.e., the number of plates with an area equal to or larger than $A$.
Most of the data except for a few largest plates follow a
linear dependence in the double log-scale of the figure. The slope of this straight line
is close to $0.25$. We first compare this sample with a pure power law (the
Pareto distribution). For this purpose we use the rank-ordering method (see
[{\it Sornette et al.}, 1996]). We put the sample into descending order
$A_1 \geq A_2 \geq ... \geq A_N$. The probability density function (PDF) of the $n$-th
rank, denoted $\phi_{n,N}(x)$, is well-known
\be
\phi_{n,N}(x) = (N-n+1)~ {N \choose n} ~F^{N-n}(x)~(1-F(x))^{n-1}~ f(x)~,
\label{mmjglw}
\ee
where $F(x), f(x)$ are the distribution function (DF) and PDF of the random values in 
question. Putting the Pareto law $F(x) = 1-(a/x)^{\mu}$, $x \geq a$, into (\ref{mmjglw}),
we get
\be
\phi_{n,N}(x) \propto \left(1 - (a/x)^{\mu}\right)^{n-1}~x^{\mu (N-n+1) -1}~.
\label{mgmjw}
\ee
The mode $M_{n,N}$ of the PDF (\ref{mgmjw}) (i.e., the maximum of the PDF) is
the most probable value of the random variable $A_n$:
\be
M_{n,N} = a \left( {N \mu +1 \over n \mu +1}\right)^{1/\mu}~.
\label{mgmmkre}
\ee
Besides, an interval around the mode containing some prescribed probability 
(say, $90\%$) can be derived from the density $\phi_{n,N}(x)$. 
The dependence (\ref{mgmmkre}) with $\mu = 1/4$ is shown as the long-dashed line
in Fig.~1. 
The two short-dashed lines
represent the upper/lower limits of the $90\%$-confidence interval. 
The data are well accounted for by the power law, 
except, perhaps, for the three smallest ranks, i.e. the three largest plates, 
the Pacific, the Antarctica, and the Africa plates, 
which fall outside the confidence interval.

\subsection{``Finite-size'' constraint on the PDF of plate sizes}

From a visual inspection of figure 1, it might be argued
that the deviation from the power law prediction (\ref{mgmmkre}) occurs
somewhat earlier, say,
at rank $n=7$, i.e., the seven
largest plates with area more than $1$
steradian belong to a different population than the rest of the plates.
The simplest possibility is that the largest plates are sensitive to 
a ``finite-size'' effect, namely that the sum of areas
over all plates must sum up to $4 \pi$. 
The largest continent and ocean plates are commensurable
with the size of the convection cells in the upper mantle and they would thus
feel the constraints due to the finite size of the Earth surface. In contrast, one
could argue that the smaller plates 
appeared as a result of the interaction and collision between
the larger plates and do not feel any superficial restriction. More precisely,
the smaller plates may be the result of fragmentation/creation processes in the
neighborhood of the borders and especially near triple points of the seven largest
plates.

The distribution $g_{N,C}(x)$ of sample values 
$x_1, ... , x_N$  conditioned by the constraint
\be
S_N = x_1 + ... + x_N = C~,
\label{mhmh}
\ee
where $C$ is a constant ($4\pi$ for the plates) is modified from 
its unconditional Pareto density expression
$f(x) = \mu~a^{\mu}/x^{1+\mu}$ for $x \geq a$ and $f(x) = 0$ for $x<a$.
The lower threshold (the minimum value) for the plate data is $a = 0.00178$ steradian
and corresponds to the smallest documented plate in [{\it Bird}, 2002].
Denoting the
unconditional density of the sum $S_k =  x_1 + ... + x_k, k = N-1, N$ by
$s_k(x)$, we have $g_{N,C}(x) = s_{N-1}(C-x)~ f(x) / s_{N}(C)$,
for $a \leq x \leq C$.
Thus, the constraint (\ref{mhmh}) decreases the unconditional Pareto density
$f(x)$ by a factor $s_{N-1}(C-x)/s_{N}(C)$ which acts as a ``taper.'' 

In order to use the Maximum Likelihood (ML) method for the estimation of the exponent $\mu$,
we need the vectorial distribution of conditional sample to 
take into account the interdependence between the different variables (areas of the plates)
 $x_1, ..., x_{42}$ induced by the constraint (\ref{mhmh}). The corresponding
likelihood function is therefore
\be
\phi(x_1, ..., x_{42} | \mu) = {\delta(x_1 + ... + x_{42} - 4\pi) \over s_{42}(4\pi)}~
f(x_1|\mu)...f(x_{42}|\mu)~.
\label{mhmhtr}
\ee
The resulting ML estimate is $\mu = 0.25 \pm 0.05$. With this value,
we generate an artificial sample of 1045
conditional 42-dimensional (42D) vectors with the
condition (\ref{mhmh}) with $C = 4\pi$. Rank-ordering each of these
1045 vectors, we determine their sample medians $M_1, ..., M_{42}$, where
$M_j$ is the median of the $j$-th rank. These conditional medians are slightly smaller 
than given by (\ref{mgmmkre}) for the unconditional Pareto distribution.
The conditional distribution (\ref{mhmhtr}) allows us
to construct a confidence domain for the 42D random vectors,
defined as a ``corridor'' of the form $[c M_j; (1/c) M_j]$, the constant $c = 0.244$
being chosen such that $95\%$ of vectors fall within this corridor. 
The medians $M_1, ..., M_{42}$ and their corridor 
are shown in Fig. 2: all samples of the tectonic plates falls within the $95\%$
confidence corridor, showing that (in contrast with the ``pure'' Pareto 
used in Fig.1) the Pareto model together with the total area constraint (\ref{mhmh})
accounts satisfactorily for all the data, including the largest plates.

We have performed many other tests which all confirm the goodness of fit
of the constrained Pareto distribution (or of 
its approximation by the truncated Pareto distribution). Using Wilks' theorem [{\it Rao}, 1965],
we used the so-called Gamma-distribution with 
complementary cumulative distribution function $G(x)$, truncated from both sides
$G(x|a,b) = C(a, b) {e^{-a x} \over x^{1+b}}$ for $u_0 \leq x \leq u_1$
and find that (1) the introduction
of the additional parameter $a$ in the exponential does not increase significantly
(probability of rejection $38\%$)
the likelihood as compared with the truncated Pareto and (2) 
the truncated Pareto distribution is 
much preferable compared with the truncated exponential distribution
(significance $1-10^{-8}$).
This conclusion is strongly confirmed by tests using so-called
sufficient statistics for both the power law and exponential models.
We also tested the popular lognormal DF, which is
a natural model for fragmentation since it is the limit distribution for processes of
subdivision of rock under rather general assumptions (basically multiplicative
processes) [{\it Kolmogorov}, 1941]. Using the Kolmogorov test and the bookstrap method,
we estimate that the data deviates significantly from the lognormal DF at the
$95\%$ significance level.

\section{Discussion}

Our main result is that Bird's suggestion [2002] of a power-like
distribution of plates areas is confirmed up to the largest plates, when taking
into account the very natural constraint that the plates areas must sum up to $4 \pi$
steradians. 

Given this strength of the validity of the Pareto distribution 
for the $42$ plates documented in [{\it Bird}, 2002], 
one can expect that it can be extrapolated
beyond this range to smaller yet undetected plates. Using the
complementary Pareto distribution $(a/x)^{\mu}$ with $a=0.00178$ steradian
and $\mu = 0.25$, this extrapolation predicts
$100$ plates larger than $1.35~~10^{-4}$ steradian
($5,500$km$^2$ or $74 \times 74$ km) [{\it Bird}, 2002]. 
The total area needed to define another 58 plates
ranging from this size up to the smallest size in PB2001 would be only about
$0.028$ steradians, which could be taken from large plates like EU and NA without
materially affecting their areas. 
As discussed by Bird [2002], the places where additional small plates
are most likely to be recognized is within the zones of distributed deformation
identified in [{\it Bird}, 2002], which have total area of $0.838$ steradians (6.7\% of Earth). 

The main suggestion inspired by Bird's discovery is that the separation of scales
between a plate tectonics involving a few major plates
at large scales and tectonic deformation at smaller scales
may be an artifact that resulted historically from insufficient statistics
and more recently from a
not quite correct interpretation of ``visual'' deviations from the pure power law.
With the improved extended compilation of [{\it Bird}, 2002] which will 
probably still improve
in the years to come, we conjecture that
there is a hierarchy of plates and block sizes from the largest ones
to the smallest scales, which may be the result of the self-organization of 
the plate motions, creation and destruction over millions of years. In other words,
we suggest that our view of plate tectonics should be revised
in terms of a dynamical model
of plates with creation, fragmentation and destruction that may occur at all scales.
This view may provide a link between the large organization of major plates
down to the block structures at scales of tens of kilometers or less discussed in
the introduction.

What could be the mechanism(s) at the origin of the observed power law
distribution of plate areas? We propose a fragmentation model with source terms,
which can be progressively
enriched by taking into account the kinematic compatibility constraints 
between plates. There is a large class of such fragmentation models
that can be solved explicitely within the scaling
theory of [{\it Cheng and Redner}, 1988]. In a nutshell, 
their approach is to write down a {\it
linear} integro-differential coupled equation for the time evolution of
fragment populations. The fragment distribution $p(l)$
is characterized by the moments $m_{\alpha} = \int_0^{\infty} l^{\alpha} p(l)
dl$. Note that calculating a moment corresponds to taking
 the Mellin transform of $p(l)$. The
moments $m_{\alpha} $ give access to $p(l)$ by taken
the inverse Mellin transform. The equations of evolution for the fragment
population then transform into the following recurrence equation for the
moments: $m_{\alpha + \beta} = \omega {1- \alpha \over L_{\alpha} - 1}
m_{\alpha}$, where $\beta$ is the exponent describing the overall rate of
breakup (assumed to be proportional to $l^{\beta}$),
$\omega$ is a normalizing constant and $L_{\alpha} = \int_0^1 x^{\alpha}
b(x) dx$,
where $b(x)$ is the probability that the fragment ratio be $x$. From the factor
${1 \over L_{\alpha} - 1}$, if
there is a value $\alpha^*$ such that $L_{\alpha^*} = 1$, then all moments with
$\alpha > \alpha^*$ will become infinite. Provided reasonable analyticity
conditions hold, it follows that the value $\alpha^*$
is a pole of $m_{\alpha}$. Taking the inverse Mellin transform of
$m_{\alpha}$ then allows us to
get $p(l)$ and, using the existence of the pole at $\alpha^*$, this
immediately predict that $p(l) \sim l^{-(1+ \beta + \alpha^*)}$ is a power
law distribution. This model was used by Ouillon et al. [1996] to
account for the power law distribution of joint spacing 
in a granitic massif in Saudi Arabia, with exponent $\mu_s=1/2$. Note that
block sizes are proportional to spacing taken to the second power, 
which then predicts $\mu = \mu_s/2 =1/4$, which is in agreement with our best estimate
$\mu =0.25$ for the plate area distribution.
This agreement between two very different systems may be fortuitous but is 
worth keeping in mind, especially in view of our claim of a continuity of the
physics of fragmention from the large plates to the blocks.
We note also that the fragmentation model of [{\it Cheng and Redner}, 1988]
predicts generically log-periodic corrections to the main power law since,
for a large variety of $b(x)$, there are
solutions of $L_{\alpha^*} = 1$ with complex exponents $\alpha^*$. This
has been shown to signal the existence of discrete scale invariance 
[{\it Ouillon et al.}. 1996].
It is tempting to interpret in this way the three approximately regular oscillations
observed in figure 1, with a prefered scaling ratio approximately equal to $10$.
This value is in good agreement with the square of the prefered scaling ratio $3.5$ 
found in the discrete hierarchy of linear sizes in fracture of rocks by
Sadovskiy et al. [1984].

Our present conclusion on a universal fractal character of plate
fragmentation does not prevent the existence of some underlying
distinctions. For instance, P. Bird (private communication and [Bird,2001])
proposes three main tectonic origins for plates:
plume-assisted rifting for the larger plates 
(ranks 1-8 in Figs.~1 and 2), continental collision stress for the
intermediate plate sizes (ranks 8-20), and back-arc spreading
for the smallest plates (ranks 21-42). In Figs.~1 and 2, one can discern 
slightly different patterns of the 
tail behavior of these three subgroups. However, 
any formal statistical analysis of these possible distinctions would
be, to our opinion, excessive due to the extreme smallness of the
data set.

\acknowledgments
We acknowledge stimulating discussions with P. Bird and Y.Y. Kagan and thank P. Bird
warmly for sharing his data.

%% ------------------------------------------------------ %%
%
%%  REFERENCES
%
%% ------------------------------------------------------ %%

   % Use "\cite" commands to call out citations,
   % and then use LaTeX's "thebibliography" environment 
   % for the reference list.  Please note that the 
   % \begin{thebibliography}{} command is followed by a 
   % null argument {} (see the example below).
   %
   % Each reference has a \bibitem command to define 
   % the citation format and the symbolic tag, as well 
   % as a \reference command which sets up formatting 
   % parameters for the reference list itself.  Authors 
   % choose what markup appears in the text, but it must 
   % match the tags used with the corresponding \bibitem 
   % command, for example, \cite{can95}
   %
   % Try to use \cite for all references in your document. 
   % Use \markcite commands for the format such as 
   % {\it Kurtz and King} [1994].
   % HOWEVER, THIS FORMAT SHOULD BE USED AS LESS AS POSSIBLE.
   %

{}
\end{article}
   % You must type an \end{article} 
   % command after the references.

\pagebreak

\begin{figure}
\includegraphics[width=\textwidth]{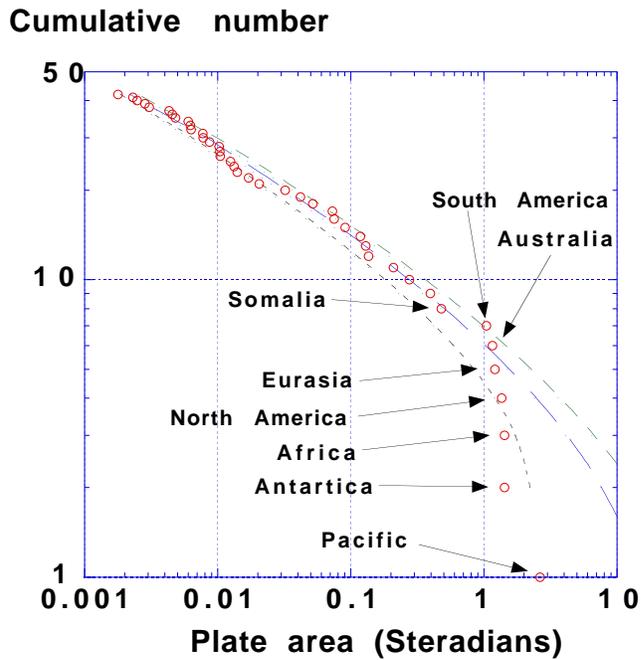}
\label{figpdfML}
\caption{Complementary cumulative distribution of 
the areas of tectonic plates (open circles)
compared to the fit with the formula 
(\ref{mgmmkre}) for a power law
(central long-dashed line) with exponent $\mu = 0.25$ and $a=0.002$. The small-dashed
line and medium-dashed line provide a confidence interval defined in the text.
}
\end{figure}

\clearpage

\begin{figure}
\includegraphics[width=\textwidth]{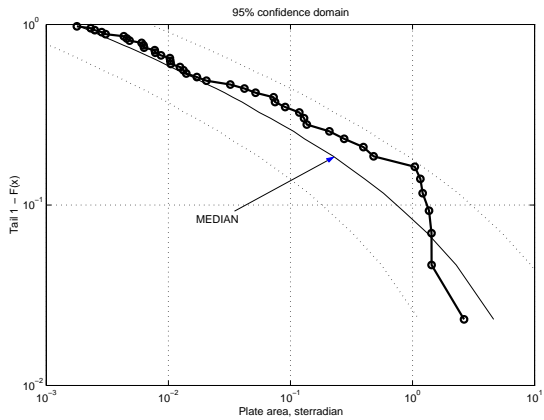}
\label{fig2}
\caption{Medians $M_1, ..., M_{42}$ (continuous line) and their corridor at
the $95\%$ confidence level
delimited by the two dotted lines of the conditional Pareto distribution
$\phi(x_1, ..., x_{42} | \mu)$ given by (\ref{mhmhtr}) compared with the 
empirical cumulative distribution
of the $42$ plate areas (circles linked by straight segments),
documented by Bird [2002].
}
\end{figure}

\end{document}